\documentclass{nature}

\usepackage[utf8]{inputenc}

\usepackage{eso-pic}
\usepackage{graphicx}
\usepackage{color}
\usepackage{amsmath}
\usepackage{caption}
\usepackage{setspace}
\usepackage{ulem}
\usepackage{placeins}
\usepackage{hyperref}

\title{AI powered, automated discovery of polymer membranes for carbon capture}

\author{Ronaldo Giro$^1$, Hsianghan Hsu$^2$, Akihiro Kishimoto$^2$, Toshiyuki Hama$^2$, Rodrigo F. Neumann$^1$, Binquan Luan$^3$, Seiji Takeda$^2$, Lisa Hamada$^2$ \& Mathias B. Steiner$^1$}

\date{May 2021}

\begin{document}

\maketitle

\begin{affiliations}
 \item IBM Research, Av. República do Chile, 330, CEP 20031-170, Rio de Janeiro, RJ, Brazil.
 \item IBM Research, 7-7, Shin-Kawasaki, Saiwai-Ku, Kawasaki, Kanagawa, 212-0032, Japan.
 \item IBM Research, 1101 Kitchawan Rd PO Box 218
Yorktown Heights, NY 10598-0218, USA.
\end{affiliations}

\begin{abstract}
The generation of molecules with Artificial Intelligence (AI) is poised to revolutionize materials discovery. Potential applications range from development of potent drugs to efficient carbon capture and separation technologies. However, existing computational frameworks lack automated training data creation and physical performance validation at meso-scale where complex properties of amorphous materials emerge. The methodological gaps have so far limited AI design to small-molecule applications. Here, we report the first automated discovery of complex materials through inverse molecular design which is informed by meso-scale target features and process figures-of-merit. We have entered the new discovery regime by computationally generating and validating hundreds of polymer candidates designed for application in post-combustion carbon dioxide filtration. Specifically, we have validated each discovery step, from training dataset creation, via graph-based generative design of optimized monomer units, to molecular dynamics simulation of gas permeation through the polymer membranes. For the latter, we have devised a Representative Elementary Volume (REV) enabling permeability simulations at about 1,000x the volume of an individual, AI-generated monomer, obtaining quantitative agreement. The discovery-to-validation time per polymer candidate is on the order of 100 hours in a standard computing environment, offering a computational screening alternative prior to lab validation.
\end{abstract}

So far, the discovery of new materials has been a time consuming and resource intensive effort. The following trial-and-error approach is typically employed: identifying known materials with properties similar to the new material's target properties and then modifying or combining them to achieve the desired outcome. The approach was driven by a specialist's knowledge, laboratory experimentation, and it could take years to yield results. The computer revolution has brought about powerful simulation techniques, such as the Density Functional Theory (DFT)\cite{Hohenberg1964,Kohn1965} method, that are aiding materials discovery today. High-Throughput Computational Materials Screening and Design (HCMSD) methods have enabled substantial speed-up of the process\cite{Zhang2012,Jain2011,Zhang2017,Mounet2018,Horton2019,Brunin2019}. However, a limitation of HCMSD is that it relies solely on time consuming \textit{ab-initio} calculations of the occurring physical and chemical processes\cite{Jain2011,Zhang2017,Horton2019,Brunin2019}. 

The emergence of repositories with large sets of experimental and simulation data has enabled the application of AI methods as a data-driven pathway to materials discovery\cite{Curtarolo2013,Ramprasad2017,Hafiz2018,Cai2020}. AI based materials design\cite{Ramprasad2017,Liu2017,Lu2017,MannodiKanakkithodi2018,Kim2018a} has a potential advantage over HCMSD: while still relying on materials screening, it does not depend solely on \textit{ab-initio} calculations. Recently, the Inverse Materials Design (IMD) method\cite{Kim2018,Wu2019} has shown its potential: An algorithm creates optimized molecular structures based on a pre-defined feature vector containing a set of materials target properties. To complete the discovery process, the IMD output would have to undergo physical validation. For polymer membranes, this validation is needed at meso-scale where the process-relevant properties of amorphous materials emerge. At that scale, however, automated \textit{ab-initio} simulation methods for validating complex materials do not yet exist.

To exemplify the issue, let us consider the case of carbon dioxide separation in post-combustion applications. From a process perspective, polymer membranes\cite{Noro2015,Firpo2019,Powell2006} have certain advantages, among them high tolerance for the challenging operating conditions and adaptability to the existing power plant steam cycle. However, a polymer's gas filtration performance cannot be derived from the physical and chemical properties of the monomer constituents alone. Rather, it is determined by the heterogeneous internal structure and complex morphology of the amorphous polymer. Therefore, predicting and validating a membrane's gas permeability remains a major challenge\cite{Kong2019}.

Encouragingly, it was recently reported that machine learning applied to known polymer repeat units can predict gas separation performance of polymers that were not previously tested for these properties\cite{Barnett2020}. However, the reported method did not offer the IMD benefits of generating optimized monomer units and, therefore, cannot generate new polymer candidates. Also, it lacked automated outcome validation of physical performance. 
In the following, we report a fully automatized, \textit{in silico} materials discovery workflow that overcomes those limitations. For demonstrating the methodological advancements with regards to the discovery of small molecules, we have applied the workflow to the generative design and physical validation of homo-polymers optimized for carbon dioxide filtration under industrial process conditions.

\begin{figure}
  \includegraphics[width=\linewidth]{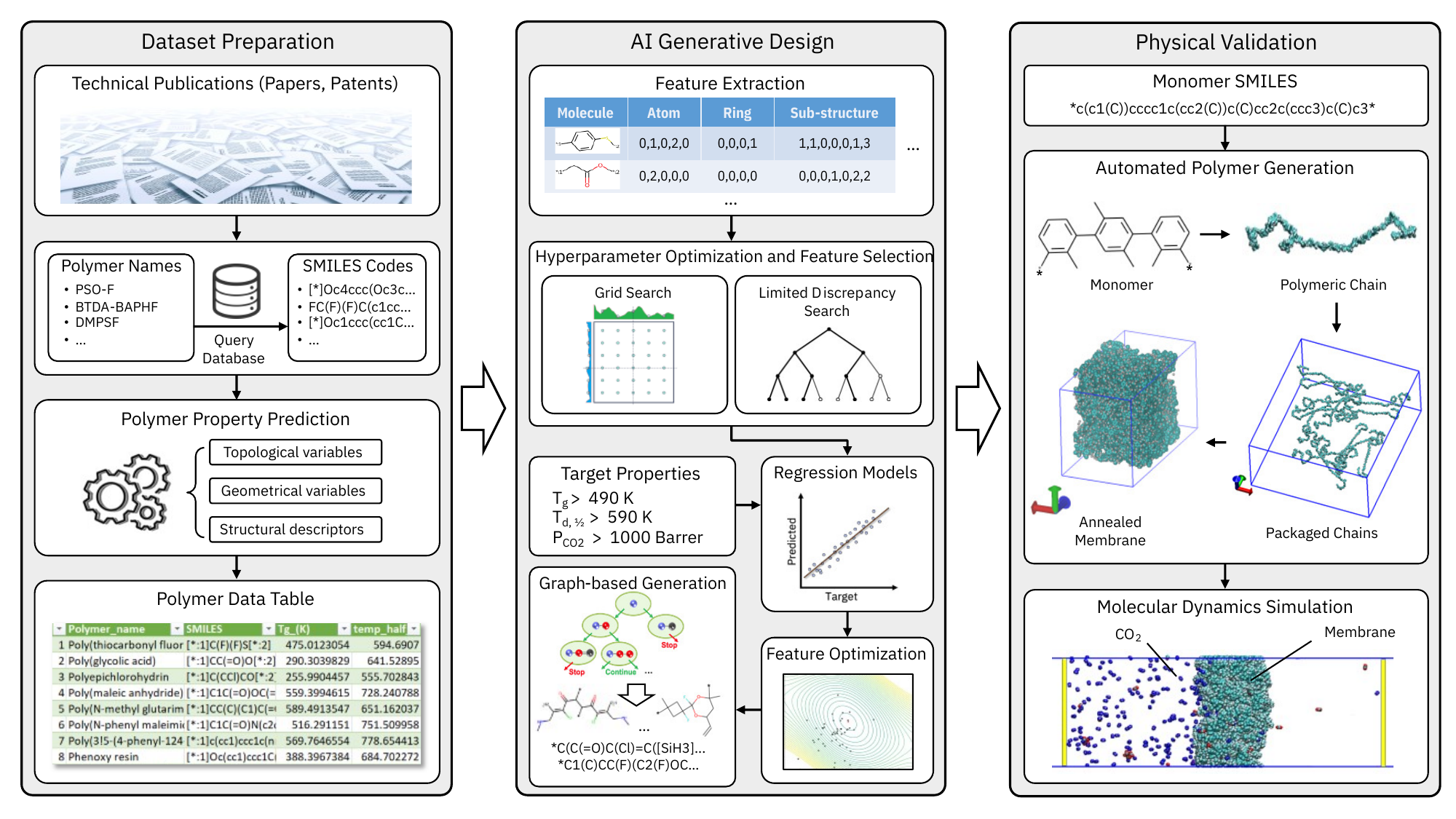}
  \caption{Automated, end-to-end computational discovery and physical validation of polymer membranes. The workflow consists of training dataset preparation, AI based generative monomer design, and physical validation of polymer membrane filtration with molecular dynamics simulations.}
  \label{fig1}
\end{figure}

In Fig.\ref{fig1}, we show the discovery workflow end-to-end, from training dataset preparation, via AI generative design to physical validation by molecular dynamics simulation. Small organic molecules, or monomer units, that typically qualify as candidate building blocks for polymer membranes, are often treated as graphs and can be converted to computer readable SMILES format\cite{Weininger1988}. For training dataset preparation based on SMILES, we have extended a Quantitative Structure–Property Relationships approach\cite{jozefbook} and made it available through our Polymer Property Prediction (PPP) engine. For AI generative modeling, we have created an Inverse Materials Design (IMD) engine\cite{Takeda2020} which extracts molecular features with regression and performs graph-based construction with SMILES input. Finally, the discovered monomers are physically validated at meso-scale by means of automated Constant Pressure Difference Molecular Dynamics (CPDMD) simulations\cite{Kong2019}, a non-equilibrium method suited for predicting a polymer membrane's gas filtration performance under realistic process conditions. The workflow is aligned with our discovery approach in reference \cite{PyzerKnapp2022}.

\begin{figure}
  \includegraphics[width=\linewidth]{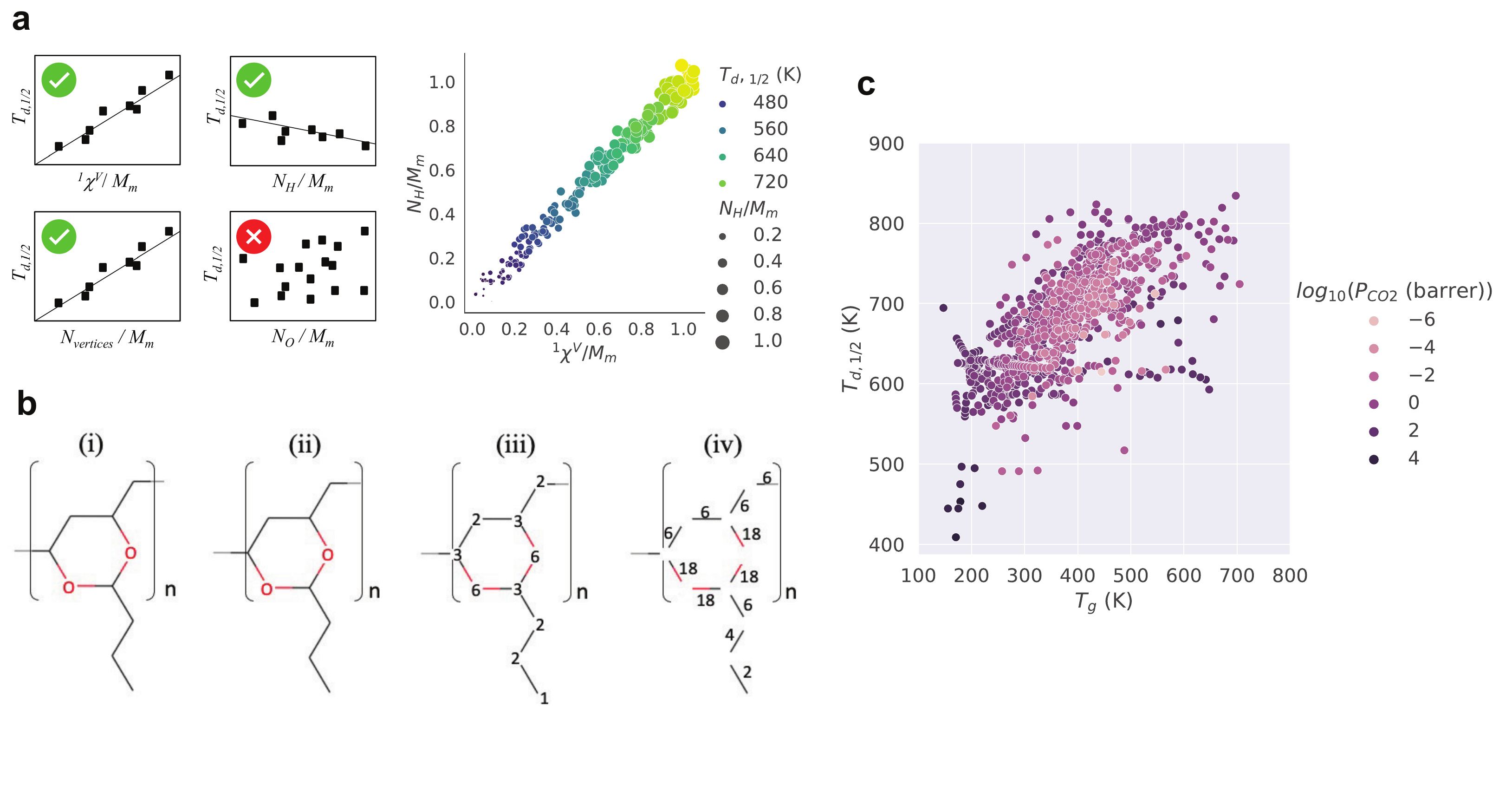}
  \caption{Dataset preparation with the Polymer Property Prediction (PPP) engine. (a) Example of half-decomposition temperature $T_{d,1/2}$ which is calculated according to Eq.\ref{eq:tdhalf} (see Methods Section). For $T_{d,1/2}$, structure-property correlations were established with the first-order (bond) connectivity index $^1\chi^V$, the number of hydrogen atoms $N_H$ and the number of vertices $N_{vertices}$ in the hydrogen-suppressed graph representation of a polymer's monomer\cite{jozefbook}. (b) Example of a hydrogen-suppressed graph representation for poly(vinyl butyral) built from the polymer name and the corresponding OPSIN SMILES string\cite{opsinwebsite}. (i) Schematic representation of poly(vinyl butyral); (ii) alternative representation with brackets not intersecting the bonds; (iii) hydrogen-suppressed version of (ii) with the valence connectivity indices $\delta^V$ in the vertices and (iv) with the bond indices $\beta^V$ in the edges, respectively (see Eqs.\ref{eq:deltav} and \ref{eq:betav} in the Methods Section). (c)  Multi-dimensional property distribution of the input dataset containing 1169 homo-polymers.}
  \label{fig2}
\end{figure}

The training dataset preparation sequence is shown in the left box of Fig.\ref{fig1}: polymer name collection from existing data sources, polymer name conversion into the OPSIN SMILES strings\cite{opsinwebsite}, and polymer name mapping to suitable target polymer properties and their respective numerical values. As high-quality lab data is often sparse or not available at all, we have used PPP for calculating polymer properties based on topological variables, such as connectivity indices, combined with geometrical variables and other structural descriptors\cite{jozefbook}. In Fig.\ref{fig2}, we illustrate the PPP conception and outline as a representative example the prediction of the half-decomposition temperature $T_{d,1/2}$ (see Methods Section). In the example of poly(vinyl butyral) shown in Fig.\ref{fig2}b, we obtain $T_{d,1/2}$ = 646 K which is in agreement with the experimental value of 645 K\cite{jozefbook}. Similarly, we have used PPP to predict the glass transition temperature $T_{g}$ (in K) and $CO_{2}$-permeability (in Barrer) for all 1169 homo-polymers in our dataset shown in Fig.\ref{fig2}c. These are suitable target properties for informing generative design of new monomers to be validated in gas separation membranes at process level.

\begin{figure}
  \centering
  \includegraphics[width=\linewidth]{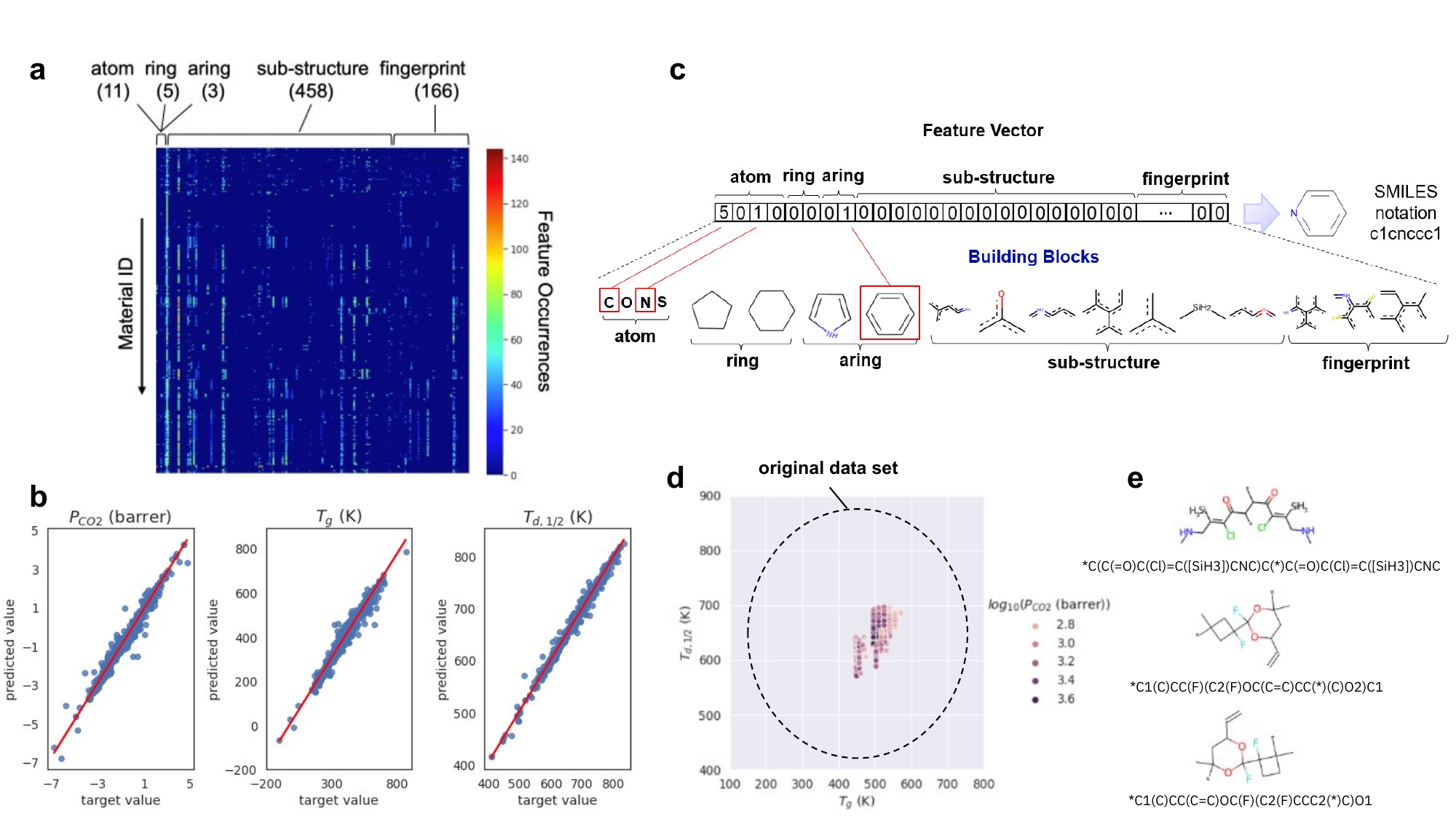}
  \caption{AI generative modeling with the Inverse Materials Design (IMD) engine. (a) The structure of each monomer in the input dataset is encoded as a feature vector. Here, the aring feature label stands for aromatic ring and the numbers under each label indicate their respective occurrences. (b) Regression results for each of the pre-defined target properties:
  $P_{CO2}$, $T_{g}$, and $T_{d,1/2}$. Blue circles: training data, red line: data fit. (c) Feature vector representation with encoded molecular building blocks: atoms, rings, aromatic rings, sub-structures, and fingerprints. Decoding the feature vector reveals a pyridine molecule for which the SMILES representation is also shown. (d) Multi-dimensional property distribution of the generated dataset. (e) AI generated monomers selected for physical validation in polymer representations via molecular dynamics simulation. The SMILES representation of each monomer is shown under the respective unit.}
\label{fig3}
\end{figure}

The AI generative design sequence is shown in the middle box of Fig.\ref{fig1}: feature extraction and selection, regression model training, feature optimization, and graph-based structure generation. For automatically generating new monomers with pre-defined target properties, we have represented each of the homo-polymers in the input dataset by its monomer in the form of a feature vector. As visualized in Fig.\ref{fig3}a, each feature vector contains structural descriptors such as the numbers of heavy atoms, rings, aromatic rings, substructures, and fingerprints. 

For molecular property prediction, we have trained and cross-validated regression models with respect to multiple sets of feature vectors and to each of the pre-defined target properties: half-decomposition temperature $T_{d,1/2}$, glass transition temperature $T_g$, and CO$_2$-permeability $P_{CO2}$ as shown in Fig.\ref{fig3}b. 

Specifically, we have trained five regression models: Lasso Regression, Ridge Regression, Elastic Net Regression, Random Forest Regression, Kernel Ridge Regression and Support Vector Regression (SVR). For training each model, we have applied both hyperparameter optimization and feature selection. For the Kernel Ridge and SVR models, respectively, we have developed a new method that efficiently performs hyperparameter optimization and feature selection simultaneously (see the Method Section). For the other models, we have performed grid search for optimizing hyperparameters while selecting features using the SelectFromModel class in Scikit-learn \cite{scikit-learn}. To maximize accuracy, we have selected the SVR model yielding the best cross validated $R^2$ score. 

For generative design, we have then optimized the feature vectors through inversion of the prediction model within the pre-defined target property ranges which were set to: 550 K $<$ $T_{d,1/2}$ $<$ 700 K; 400 K $<$ $T_g$ $<$ 600 K and 630 barrer $<$ $P_{CO2}$ $<$ 4000 barrer. We have expanded the optimized feature vectors to molecular structures through an advanced version of the Molecular-Customized McKay’s Canonical Construction Path Algorithm\cite{Takeda2020,Takeda2022}.
The algorithm repeats cycles of connecting structural fragments such as atoms, rings, and substructures, and cycles of feature screening. Our methodological advancements (see Method Section and Supplemental Information) enable the application of graph-based generative design to complex molecular structures. In addition, the IMD allows for defining design rules with regards to structural constraints, the range of the number of substructures, as well as fragment patterns. As a result, chemical subject matter expertise can inform the generative design process. In Fig.\ref{fig3}c, we have visualized an example of how the generative algorithm transforms a feature vector into a molecular structure. A feature vector encodes structure-specific information for each molecular building block. The algorithm generates a specific molecular expression of the feature vector based on a library of building blocks created during the feature vector encoding process.

After completing the AI generative design sequence and screening our initial discovery results for target property range and discrepancies between predicted and calculated polymer property values, we have obtained a set of 784 new monomer candidates shown in Fig.\ref{fig3}d. In the following, we will physically validate the most promising of the discovered monomers, visualized in Fig.\ref{fig3}e, in a polymer membrane configuration by means of automated molecular dynamics simulation.

\begin{figure}
  \centering
  \includegraphics[width=14cm]{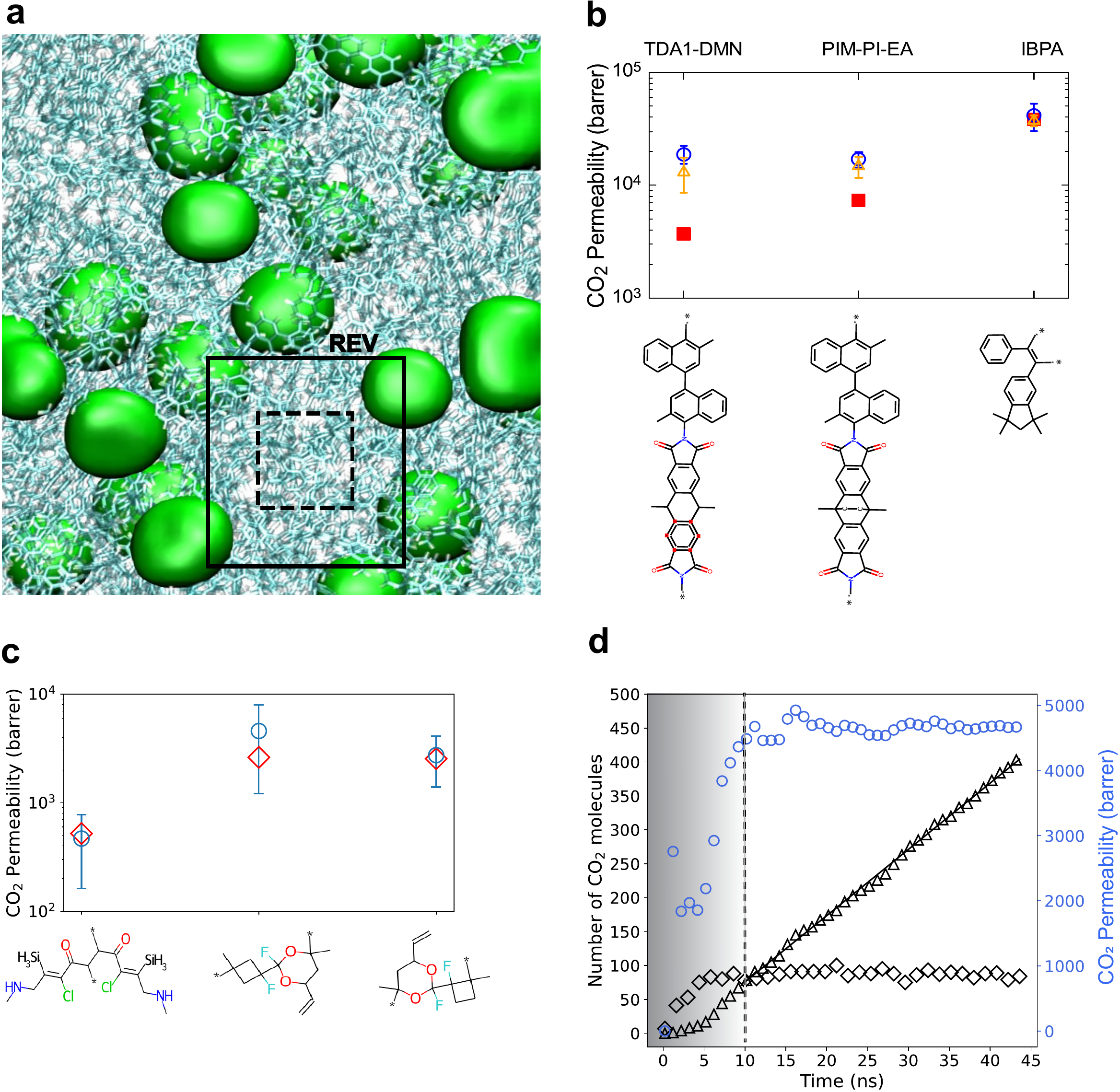}
  \caption{Physical validation of AI discovered polymers with automatized Constant Pressure Difference Molecular Dynamics (CPDMD) Simulations. (a) Cross-sectional view of a meso-scopic polymer membrane representation. The volumes rendered in green color are void spaces visualized using 3V\cite{Voss2010,3vwebsite}. The solid frame visualizes the Representative Elementary Volume (REV) concept; the dashed frame does not adequately capture the membrane's average porosity. (b) Representative monomers for determining the polymer REV. The asterisks indicate the head and tail atoms, respectively. Carbon atoms: black; oxygen atoms: red; nitrogen atoms: blue. CPDMD simulation results are shown as blue circles: 20,000 atoms; orange triangles: 40,000 atoms. Experimental reference: red squares. (c) CO$_2$ permeability of representative, AI designed polymers. Red diamonds: AI model predictions; blue circles: CPDMD simulation results. (d) CPDMD simulation of CO$_2$ filtration dynamics. CO$_2$-permeability: blue circles; number of CO$_2$ molecules permeated: black triangles. The line is a linear fit to the data. Number of CO$_2$ molecules sorbed in the membrane: black diamonds.The shaded area indicates the transient simulation regime.}
\label{fig4}
\end{figure}

The automated molecular dynamics simulation sequence is shown in the right box of Fig.\ref{fig1}: creation of SMILES representation of the discovered monomer, creation of a polymer membrane representation with the discovered monomer, and physical simulation of the gas filtration process through the membrane. Prior to applying the above sequence to the newly generated monomers, we have confirmed the suitability of the CPDMD method for physical validation of membrane performance through extensive benchmark analyses with known polymers, (see Methods Section and Supplemental Information). A fundamental question occurs with regards to the minimum volume that adequately represents the properties of complex materials at mesoscopic scales. In the present case of gas separation with polymer membranes, we have adopted the concept of Representative Elementary Volume (REV), which is routinely used for characterising porous media\cite{CostanzaRobinson2011}. REV can be understood as the smallest material volume for which a physical property can be determined such that it yields a value that is representative of the bulk. To illustrate this concept, we show in Fig.\ref{fig4}a cross sections through computational membrane representations exhibiting porosity variations. Depending on the region sampled, a material's porosity can be smaller or larger than the bulk average. If probed at or above REV-level, the porosity value matches the bulk average.  

To probe REV with regards to our polymer permeability simulations, we have investigated three representative polymers with relatively high permeability-values: TDA1-DMN, PIM-PI-EA and IBPA shown in Fig.\ref{fig4}b - bottom of figure from left to right. For each of the three polymers, we have performed five independent CPDMD simulations using the simulation box set up shown in Fig.\ref{fig1}. By doubling the number of atoms in the simulation and keeping the membrane thickness fixed at 6nm, the cross-sectional area also doubled, from around 50 nm$^2$ to 100 nm$^2$. By probing larger areas and randomly sampling the amorphous polymeric chains making up the membrane, we observe a trend in Fig.\ref{fig4}b that the simulations with larger volume tend to approach the experimental values.

After establishing both the automated CPDMD simulation protocol and the REV determination (see Methods Section and Supplemental Information), we have proceeded with the validation of the three shortlisted new polymer candidates generated by IMD. Two similar monomers were selected for investigating how head and tail positions influence the simulation results. For each of the three polymers, we have performed five independent CPDMD simulations using the simulation box in Fig.\ref{fig1}. As a key result of our investigation, we show in Fig.\ref{fig4}c the simulated CO$_2$ permeability values obtained for the polymer membrane representations of AI designed monomers. We observe quantitative agreement, within the error bars, of AI predicted permeability values and the CPDMD based physical validation results. To our knowledge, this is the first computational performance validation of an AI discovered, complex material.

For analyzing the filtration dynamics, we show as a representative example for one of the polymers in Fig.\ref{fig4}d the simulated CO$_2$ permeability along with the number of CO$_2$-molecules permeated and sorbed, respectively, as function of simulation time. The corresponding CO$_2$ density profile evolution is shown in Supplemental Fig.S8. Initially, the CO$_2$ molecules are located at the left-hand side of the membrane, see simulation box in Figs.\ref{fig1}, which is connected to the gas feed chamber. As shown in Fig.\ref{fig4}d, CO$_2$ molecules are penetrating the membrane at 1 ns. The membrane saturation level of roughly 100 CO$_2$ molecules on average, corresponding to an average density of 0.06 - 0.07 g/cm$^3$, is reached at about 5ns. At 10 ns, the CO$_2$ permeability has converged towards the saturation value of 5000 barrer. After 30ns, the permeability fluctuations have disappeared and the membrane filtration has reached a steady state. The steady-state filtration regime is characterized by the constant slope of permeated CO$_2$ molecules as function of time from which the permeability value can be extracted using eq. \ref{eq:2} (see Methods Section).

The main transport characteristics captured by the CPDMD simulations are (i) the interaction between gas molecules and polymer membrane which determines solubility and, consequently, the selectivity with regards to a specific gas molecule and (ii) the diffusion of gas molecules through the void spaces between packaged polymer chains which determines the membrane's permeability. We note that while the thickness of manufactured polymer membranes are typically on the order of micrometers, the much thinner simulated membrane predicts experimental permeability values reasonably well, within the same order of magnitude.  As expected, the accurate and repeatable determination of gas permeability is limited experimentally as well as theoretically and large error margins are an intrinsic characteristic associated with the properties of amorphous materials\cite{Kong2019}.

Using a standard computational framework (one Intel Xeon E5-2667 CPU, one NVIDIA Tesla K80 GPU), the overall computation time, from dataset preparation to AI generation to physical validation, is of the order of 200 hrs for polymeric membranes with higher permeabilities (above 1000 barrer). A computational bottleneck currently exists for reaching gas saturation and steady state filtration in lower-permeability membranes as shown in Supplemental Fig.\ref{supfig8}. 

Future extensions of this work would benefit from advanced representations of a membrane's morphology. One example would be to pack the monomers randomly in a virtual cubic box and connect their head and tail atoms according to predefined probabilities - instead of packing the polymer chains randomly. This would more closely resemble the actual polymer formation process. In generative modeling, the extension to molecular structures with higher complexity and the introduction of a user-defined objective function could open the pathway to the generation of polymers with higher complexity, such as block co-polymers. We expect that adding target properties for molecular selectivity to the optimization workflow and extending the generative algorithms to the design of co-polymers will further improve discovery outcomes.

In summary, we have reported fully automated, end-to-end computational discovery of polymer membranes for carbon dioxide separation. We have demonstrated each discovery step, from automated training data and feature vector creation via generative inverse design of new monomers to non-equilibrium molecular dynamics simulation of gas filtration by the polymer membrane. Molecular dynamics simulations successfully predict a polymer's filtration dynamics and permeability if performed with a minimum representative volume of the complex material. For computationally designed polymers, we have obtained quantitative agreement between the CO$_2$-permeability predictions by means of the AI models and the molecular-dynamics based, physical process simulations. Our work opens a pathway for advancing AI generative design beyond small-molecule applications and will substantially accelerate the discovery of complex materials for scaled applications. 

\begin{methods}

\subsection{Polymer Property Calculation for Automated Training Dataset Generation}

For creating the training dataset, we have collected representative homo-polymers names in IUPAC nomenclature standard, from multiple polymer classes\cite{polyinfowebsite}. We have then converted their individual monomer unit names to SMILES format (with their head and tail units tagged) using the Open Parser for Systematic IUPAC nomenclature - OPSIN\cite{opsinwebsite}. Based on our analysis of the gas separation process, we have selected three suitable figures-of-merits or target properties for polymer membranes: glass transition temperature ($\displaystyle T_g$ in K), half-decomposition temperature ($\displaystyle T_{d,1/2}$ in K), and permeability ($\displaystyle P$) for $\displaystyle {CO_2}$ (in Barrer). $\displaystyle T_g$ is the temperature above which segmental motions of polymer chains occur such that they negatively affect a polymer membrane's mechanical stability. $\displaystyle T_g$ also defines the transition limit between glassy and rubbery polymers (temperature below and above $\displaystyle T_g$, respectively). Glassy polymers dominate the Roberson upper bound\cite{Robeson2008,Robeson2015} due to higher solubility coefficient\cite{Robeson2015}, or, in other words, better selectivity. However, rubbery polymers have lower solubility and higher diffusion coefficients\cite{Robeson2015}, i.e higher permeability and lower selectivity. Similarly, $\displaystyle T_{d,1/2}$ defined as the temperature at which the loss of weight during pyrolysis (at a constant rate of temperature rise) reaches 50 percent of its final value should be reasonably high as it is a measure for chemical stability. A high permeability for $\displaystyle {CO_2}$ is desirable as a measure of the gas flux through the membrane. However, it is limited by a trade-off with the membrane's selectivity $\displaystyle P_{CO_2}/P_{N_2}$. For creating the training dataset, we have collected literature data for $\displaystyle P_{CO_2}$ and combined it with calculated data for $\displaystyle T_{d,1/2}$ and $\displaystyle T_g$.

Calculation of $\displaystyle T_{d,1/2}$:

Best structure-property correlations were established with first-order (bond) connectivity index $^1\chi^V$; number of hydrogen atoms $N_H$ and number of vertices $N_{vertices}$ in the hydrogen-suppressed graph representation of a polymer's monomer\cite{jozefbook}. The functional relation for $T_{d,1/2}$ was obtained through a linear regression against the best correlation descriptors:

\begin{equation} \label{eq:tdhalf}
T_{d,1/2}=1000((7.17N_{vertices}-2.31N_H+12.52~^1\chi^V)/M_m)
\end{equation}

Fig.\ref{fig2}b displays the calculation steps performed by the PPPE engine for poly(vinyl butyral). Starting with the (i) hydrogen-suppressed graph representation of poly(vinyl butyral) monomer and its (ii) alternative representation with the square brackets not intersecting the bonds, the (iii) valence connectivity indices $\delta^V$ in the vertices and the (iv) bond indices $\beta^V$ in the edges are calculated according to eq.\ref{eq:deltav} and \ref{eq:betav}, respectively:

\begin{equation} \label{eq:deltav}
\delta^V=\frac{Z^V-N_H}{Z-Z^V-1}
\end{equation}

\begin{equation} \label{eq:betav}
\beta^V_{ij}=\delta^V_i\delta^V_j
\end{equation}

\noindent where $Z^V$ is the number of valence electrons of an atom, $N_H$ is the number of hydrogen atoms bonded to it, and $Z$ is its atomic number. $\beta^V_{ij}$ is the product of $\delta^V$ at the two vertices ($i$ and $j$) which define a given edge or bond.

The first-order (bond) connectivity index $^1\chi^V$ of the entire molecule is defined through the summation over the edges of the hydrogen-suppressed graph:

\begin{equation} \label{eq:chi1v}
^1\chi^V=\sum_{edges}\frac{1}{\sqrt{\beta^V}}
\end{equation}

By combining Eq.\ref{eq:tdhalf} and Eq.\ref{eq:chi1v}, counting the number of vertices and the hydrogen atoms and calculating the molar mass of poly(vinyl butyral), we obtain $T_{d,1/2}$=646K which is in agreement with the experimental value of 645K\cite{jozefbook}.

\subsection{Hyperparameter Optimization and Limited Discrepancy Search}

The procedure referred to as feature selection identifies a subset of features for achieving accurate predictions, rather than using the entire set of the original features\cite{Chandrashekar2014}. In other words, feature selection allows a machine learning algorithm to learn a model in a lower-dimensional space. The dimensionality reduction typically leads to computational performance enhancements.

Hyperparameter optimization (HPO) is also key for enhancing the model performance. There are many HPO algorithms available in the literature, including grid search and Bayesian Optimization, see for example, reference\cite{Yang2020}.
In theory, hyperparameter configurations are specific to a feature set used to train a machine learning model. One set of hyperparameter configurations that works well for one feature set might not be the best for another feature set. On the other hand, both feature selection and HPO typically require intensive computation. For example, given \textit{N} features, finding an optimal feature requires ($\displaystyle 2^N$) possible feature sets. For \textit{M} hyperparameters, each of which has \textit{b} configurations after its possible values are discretized, there are ($\displaystyle b^M$) possible choices for the hyperparameter configurations. An optimal feature set and hyperparameter configurations need to be found out of ($\displaystyle 2^Nb^M$) combinations. In practice, feature selection and HPO are performed separately to reduce the computational overhead, e.g., perform HPO after feature selection selects an optimized feature set with default hyperparameter configurations. However, this approach might not represent a good combination of the feature set and hyperparameter configurations.

We have optimized an average ($\displaystyle R^2$) score of the three-fold cross validation with 10 repeats.To that end, we have developed a systematic local search algorithm that simultaneously performs feature selection and HPO for a non-linear machine learning model. This approach leads to an optimized hyperparameter configuration specific to a selected feature set. To reduce the computational overhead, our approach focuses only on small, promising search spaces where optimized solutions are likely to occur. 
We discretize possible values for each hyperparameter and formulate feature selection and HPO as a variable-value assignment task. This means that each variable corresponds to another variable to which one value needs to be assigned. The variable for a feature is set to either true or false, while the variable for each hyperparameter is set to one of the discretized hyperparameter-values.

Our approach is based on limited discrepancy search (LDS)\cite{Harvey1995,Korf1996}. The idea behind LDS has been studied in the artificial intelligence community and has a variety of applications such as\cite{Takeda2020}. LDS starts with an initial solution, i.e., initial variable-value assignment, and keeps refining it until a satisfactory solution is obtained.

Our current implementation calculates the initial solution passed to LDS as follows: It first calculates optimized hyperparameter configurations based on grid search with the whole feature set. With these hyperparameter configurations, it then computes the initial feature set based on so-called Sequential Backward Selection (SBS)\cite{Chandrashekar2014}. Our SBS implementation starts with the whole feature set. It repeats a greedy elimination of one feature (without which a score is improved) until no further improvement is obtained, .

The solution refinement step of LDS consists of a series of local search controlled by the notion of \textit{discrepancy}. Given the current best solution \textit{bs}, LDS assumes that a better solution exists in a search space whose solutions are similar to \textit{bs}. In our implementation, the discrepancy for a solution \textit{s} is defined as the number of variables whose assigned values have differences between \textit{bs} and \textit{s}. A smaller discrepancy indicates that \textit{s} is more similar to \textit{bs}.

LDS introduces a discrepancy threshold \textit{d} and performs local search in an iterative manner. After setting \textit{bs} to the initial solution calculated by SBS, LDS performs depth-first search with \textit{d}=1 and attempts to find a better solution than \textit{bs} in a search space where solutions are located that have a different value than \textit{bs} only for one variable. If no better solution is found, LDS increments \textit{d} and performs local search with \textit{d}=2. If no better solution is found again, LDS performs search with \textit{d}=3, and so on. If a better solution is found, LDS resets \textit{d}=1 and \textit{bs} to the better solution and restarts a local search with \textit{d}=1. LDS repeats these steps until the allocated time is used up or \textit{d} reaches a preset, maximum value.

There are several implementation choices for LDS to select a next variable for updating its value. Before performing a new iteration of local search, our current implementation orders variables in ascending order of the following formula: $\displaystyle w_1v$(\textit{x}) + $\displaystyle w_2u$(\textit{x}), where $\displaystyle w_1$ and $\displaystyle w_2$ are constants, $\displaystyle v$(\textit{x}) is the number of times variable \textit{x} is selected in local search, and $\displaystyle u$(\textit{x}) is the number of times variable \textit{x} fails to improve \textit{bs}. This formula attempts to remain the values of the variables unchanged that have contributed to improving a score as well as to prioritize the variables that have not been explored sufficiently.For the purpose of this study, we have chosen $\displaystyle w_1$=2 and $\displaystyle w_2$=1. 

In Supplemental Fig.\ref{supfig1}, we show a comparison of regression results obtained with and without the application of hyperparameter optimization.

\subsection{Feature Vector Optimization}

Based on graph theory and atomic configurations, there exist multiple feature types which can be combined for application of machine learning models, among them the number of heavy atoms, number of rings, substructures, fingerprints, Coulomb matrix, dipole moment, potential energy and experimental conditions\cite{Takeda2020}.

By using Eq.\ref{eq:1}, we estimate feature vector values fv based on a target property value $\displaystyle v_p$ and a regression model $\displaystyle f_p$ by minimizing the score of each feature vector $\displaystyle v$. More specifically, the minimization is performed over the square error of the estimated value which is normalized by the prediction variance $\displaystyle \sigma_p^2$ to which a penalty function is added to account for violations of structural constraints. The violation of structural constraints is evaluated by means of the realizability of a molecular structure connected by sub-structures in the corresponding feature vector:

\begin{equation} \label{eq:1}
\newcommand{\argmin}{\mathop{\rm arg~min}\limits}
{\rm fv} = \argmin_{v \in I^n} \{ \frac{|v_p - f_p(v)|^2}{\sigma_p^2} + {\rm violation}(v) \}
\end{equation}

\subsection{Generative Molecular Design}

The Molecular-Customized McKay’s Canonical Construction Path Algorithm creates molecular structures efficiently, exhaustively, and without isomorphic duplication, i.e., edge relations are preserved. A simplified version of the algorithm with an idealized construction example is visualized in Fig.S2a. Based on the root molecule graph, one vertex is extended from each orbit of the automorphism group. The graph is grown by performing a generation step to add a new vertex to extendable vertices of an existing graph, starting from an initial single vertex. At each generation step, a canonical labeling step of the current graph is performed. The labeling algorithm assigns ordinals to all the vertices of the isomorphic graphs, providing a unique vertex addition or construction order for obtaining the graph. If a new vertex coincides with the last vertex in the construction order, the generation step continues. Otherwise, the generation step terminates - it is pruned -  as its construction path generates a duplicate, isomorphic graph. Note, that the vertex here corresponds to a molecular structure and that adding a new vertex means adding an atom or a sub-structure in this application.

The advanced version of the generative algorithm \cite{Hama2020,Takeda2022} inherits user-customized design constrains such as, for example, expected or unexpected sub-structures in SMILES format, and the inverse designed feature vectors such as, for example, the number of heavy atoms, rings, and occurrences of fragment structures, and then converts them into molecular structures. Constraint functions capture design rules such as, for example, disallowing triple bonds between carbon atoms, limiting the number of molecular rings in the structure to between 4 and 9, or including preferential molecular substructures. For the purpose of this study, all constraints have been merged with the extracted feature vectors and best regression models for subsequent iterations of optimized structure generation.

An example with a ring of six atoms is shown in Supplemental Fig.\ref{supfig2}b. In a first step, the orbits of the automorphism group are obtained from the SMILES representation of a given sub-structure. We then create the isomorphic equivalent graph by replacing the atom name with the SMILES name and the minimum index number (indices 1 and 3). In this step, those vertices without "free hand" are eliminated which helps identifying the symmetry of the graph. For better handling, the isomorphic equivalent graph is further simplified to a single vertex representation by selecting vertices with minimum index number in each orbit, whereas other vertices are replaced by dummy atoms. Finally, we obtain the orbits of the automorphism group and the minimum index number of each orbit is selected to be an extending vertex of the sub-structure.

Supplemental Fig.\ref{supfig2}c shows a construction example. During the generation of a molecular structure as a colored graph (graph of various atoms) and by adding a vertex with a connecting edge one by one, the algorithm minimizes the number of vertices in order to improve the performance of the canonical labeling which is a bottleneck routine of the process. In the root graph, an extending vertex which has a minimum label in an orbit of an automorphism is considered to be a single vertex graph. In order to extend the vertices, it is replaced by an isomorphic equivalent representation. The new vertex is extended and canonical labeling of the entire graph is performed. Once the canonical construction path is validated, the original representation will be recovered. The new structure will be tested against the pre-defined design constraints. The cycle repeats until it fulfills pre-set requirements such as number of generated results with pre-defined target property values.

\subsection{Computational Representation of Polymer Membrane}

For the physical validation of AI predicted CO$_2$-permeability, we have created a method to automatically design a polymer membrane representation which is suitable for molecular dynamics simulation, see right box in Fig.\ref{fig1}. In a first step, the SMILES strings of AI designed monomers are indexed to indicate the position of head and tail atoms so they can be used as input for PySIMM\cite{Fortunato2017,pysimmwebsite}. We have then used the Force Field Assisted Linear Self-Avoiding Random Walk application in PySIMM\cite{Fortunato2017} to build a linear polymer chain with a maximum number of about 800 heavy atoms which are defined as atoms other than hydrogen. This way, we have kept the length of the polymer chain rather constant, independent of the monomer size. For describing the interactions between intra-chains and inter-chains atoms, we have used the DREIDING force field\cite{Mayo1990}. 

Once the chain building step is completed, PySIMM saves the LAMMPS\cite{Plimpton1995} topology file with the associated force field parameters. Then, the polymer chains are packaged in a 3D box using Packmol\cite{Martnez2009}. The 3D simulation box is periodic in x, y, z directions. We are aware of the limited accuracy of applying force-field parameters generated automatically by PySIMM for polymer modeling, and opls-aa parameters\cite{Jorgensen1996} can be adopted for an improved accuracy. For defining the membrane thickness, the z dimension of the box is set to 6 nm. The dimensions of the box in x and y are defined by a multiplication factor of the polymer chain size. The number of polymer chains is defined by the total number of atoms in the polymer membrane - 20,000 in the present case. To keep the membrane thickness in z-direction fixed at 6nm, rigid walls are placed in the x, y membrane planes. To avoid interactions between periodic images in z-direction, a vacuum layer with a thickness of 5 nm is placed at each side of the polymer membrane. 

The system then undergoes an equilibration process that consists of a nine-step compression-relaxation sequence, similar to the approach in reference\cite{Kong2019}: (1) energy minimization with isothermal and isochoric (NVT) MD simulation at 1 K for 100 ps, (2) NVT MD simulation at 300 K for 100 ps, (3) isothermal and isobaric (NPT) MD simulation at 300 K and 1 atm for 100 ps, (4) NPT MD simulation at 300 K and from 1 atm to 3000 atm for 100 ps, (5) NPT MD simulation at 300 K and 3000 atm for 300 ps, (6) NVT MD simulation at 800 K for 100 ps, (7) NVT MD simulation at 300 K for 100 ps, (8) NPT MD simulation at 300 K and 1000 atm for 300 ps, the steps (6)-(8) repeats 30 times, and (9) NPT MD simulation at 300 K and 1 atm for 10,000 ps. 

To account for long-range electrostatic interactions, we have adopted the reciprocal space
Particle-Particle Particle-Mesh (PPPM) method. For all calculations, we have used 1 fs time steps and a
cutoff radius of 1.4 nm for van der Waals and Coulomb interactions, respectively. To control temperature and pressure, we have used Nose-Hoover thermostats and barostats with a relaxation time of 0.1 ps and 1 ps, respectively. 

All MD simulations were carried out with the LAMMPS package\cite{Plimpton1995,Brown11,Brown12,Brown13}.
For further information regarding the effects of chosen force fields, chain lengths, membrane thicknesses and the equilibration process protocol, see Supplementary Information Figures \ref{supfig4} and \ref{supfig5}.

\subsection{Automated Membrane Validation with Molecular Dynamics Simulation}
Two types of Molecular Dynamics (MD) simulations methods have been used to investigate transport through membranes: Equilibrium MD (EMD) and Non-Equilibrium MD (NEMD). NEMD is ideally suited to represent an experimental membrane system in which an external driving force, such as a chemical potential or pressure gradient, is applied to the membrane. Specifically, we have chosen CPDMD to evaluate membrane based gas filtration\cite{Kong2019}. 

For benchmarking purpose, as shown in Supplemental Fig.\ref{supfig3}, we have chosen representative homo-polymers covering a broad CO$_2$-permeability range. For six of these homo-polymers, we have performed five independent CPDMD simulations each using the simulation box set up in Fig.\ref{fig1}. The results are shown in Supplemental Fig.\ref{supfig6}. Overall, we obtain reasonable agreement with literature values for BZ-CF3, IBPA, PIM-PI-EA and PEO, despite the large error bars for BZ-CF3 and PEO. The simulated CO$_2$-permeabilities of TDA1-DM and PI-5 are higher than the literature values, however, one of the PI-5 samples is close to the experimental value. We note that due to the amorphous nature of polymers, both experimental and simulations results typically exhibit large error bars\cite{Kong2019}.

To set up a CPDMD simulation, we have placed the membrane at the center of the simulation box with a fixed, rigid wall at each side of the membrance, 10 nm away from its surface, as shown in Fig.\ref{fig1}. To avoid interactions with periodic images in z-direction, we have placed a 5 nm vacuum layer beyond each rigid wall. The carbon atoms in the 5 \AA~ surface layer of the membrane were fixed in z-direction by a harmonic potential with a force constant of 5.0 Kcal/mol \AA$^2$. Following\cite{Kong2019}, we have estimated the number of CO$_2$ molecules in the feed chamber using the ideal gas law $N_{CO2}=N_ApV/RT$, where $N_A$ is the Avogadro's constant, $R$ is the gas constant, $p$ is the pressure set to 10 atm, $T$ is the temperature set to 300 K, and $V$ is the feed chamber volume, see Fig.\ref{fig1}. We have then performed NVT MD simulations at 300 K. Due to the pressure gradient, CO$_2$ molecules are absorbed within the membrane and, subsequently, transported to the permeate side. To maintain the same initial pressure gradient of 10 atm, we have added CO$_2$ molecules into the feed chamber while removing the molecules at the permeate side to produce a pseudo vacuum. We have run the addition/removal processes in cycles with a time interval of 200 ps following\cite{Liu2019}. We have used the DREIDING force field\cite{Mayo1990} for describing the interactions between intra-chains and inter-chains atoms. For CO$_2$ molecules, we have used the rigid model TraPPE force field\cite{Potoff2001}. For the CO2/polymer LJ interactions, we have applied the Lorentz–Berthelot mixing rules. All MD simulations were performed with the LAMMPS package\cite{Plimpton1995,Brown11,Brown12,Brown13} using the same parameters described in the previous Methods subsection.

From the $N_{CO2}-t$ slope, the permeability $P_{CO2}$ can be estimated following

\begin{equation} \label{eq:2}
P_{CO2}=\frac{(\Delta N_{CO2}/N_A)l}{A\Delta t p}
\end{equation}

\noindent where $\Delta N_{CO2}$ is the number of CO$_2$ molecules permeated within time duration $\Delta t$, $N_A$ is Avogadro's constant, $l$ and $A$ are the membrane thickness and area, respectively, and $p$ is the partial pressure - 10 atm in this case - in the feed chamber.

The termination criterion for CPDMD simulations is discussed in detail in the Supplementary Information and shown in Supplemental Fig.\ref{supfig7}. The evolution of the simulated CO$_2$ density profile across a polymer membrane is shown in Supplemental Fig.S8, complementing the simulation results shown in Fig.\ref{fig4}d for the same polymer.
\end{methods}

\begin{addendum}
 \item[Acknowledgements] We acknowledge discussion with and support by Manuela F. B. Rodriguez, Rong Chang, Daiju Nakano and Bruno Flach (all IBM Research).

 \item[Correspondence]
Correspondence and requests for materials should be addressed to mathiast@br.ibm.com

\item[Supplementary Information]
Supplementary Information, including Supplementary Table 1 and Supplementary Figures S1-S8, are available as a pdf-file

\item[Code Availability - 1]

The Polymer Property Prediction (PPP) Engine is available at

\noindent
\url{https://github.com/IBM/polymer\_property\_prediction}

\item[Code Availability - 2]

The jupyter notebook for polymer property predictions based on SMILES input is available - under doi:10.24435/materialscloud:ma-qn - at

\noindent
\url{https://archive.materialscloud.org/record/2022.99}

\item[Data Availability - 1]

The training dataset containing polymer candidates in SMILES format is available - under doi:10.24435/materialscloud:ma-qn - at

\noindent
\url{https://archive.materialscloud.org/record/2022.99}

\item[Data Availability - 2]

The dataset containing AI discovered polymer candidates in SMILES format is available - under doi:10.24435/materialscloud:ma-qn - at

\noindent
\url{https://archive.materialscloud.org/record/2022.99}

\end{addendum}

\clearpage
\newcommand{\beginsupplement}{%
        \setcounter{table}{0}
        \renewcommand{\thetable}{S\arabic{table}}%
        \setcounter{figure}{0}
        \renewcommand{\thefigure}{S\arabic{figure}}%
     }
\section*{SUPPLEMENTARY INFORMATION}
\beginsupplement

\subsection{Hyperparameter Optimization and Limited Discrepancy Search}

\begin{figure}[h]
  \includegraphics[width=\linewidth]{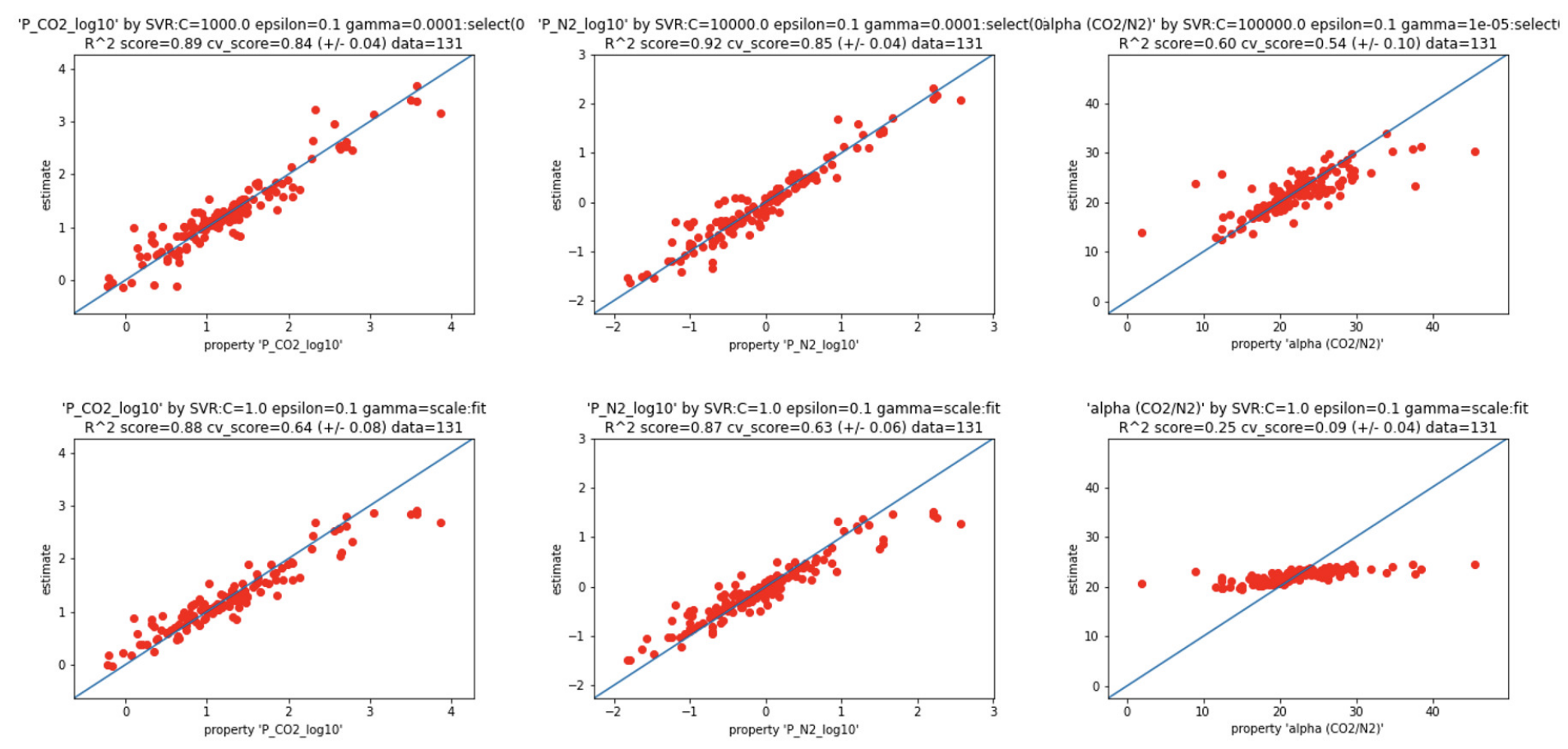}
  \caption{Comparison of regression results obtained with (upper panel) and without (lower panel) the application of hyperparameter optimization.}
  \label{supfig1}
\end{figure}

We provide supplementary data, shown in Fig.\ref{supfig1}, for supporting the discussion in the Methods Section of the main manuscript.

\subsection{Generative Molecular Design}

We provide supplementary graphics material, shown in Fig.\ref{supfig2}, for supporting the discussion in the Methods Section of the main manuscript.

\subsection{Constant Pressure Difference Molecular Dynamics Protocol}

In this section, we discuss a series of studies that we have conducted for establishing the computational protocol outlined in the Methods Section of the main manuscript with regards to the Constant Pressure Difference Molecular Dynamics (CPDMD) simulations. This includes the determination of the equilibration process to obtain the morphology of the polymer membrane, the determination of the force field, the length of polymeric chain and, finally, the thickness of polymer membrane. For our benchmark studies we have chosen eight representative homo-polymers, shown in Fig.\ref{supfig3}, covering a broad CO$_2$-permeability range .

In a first step, we have studied the effect of force field on CO$_2$-permeability. The atomic interactions in the polymer membranes are described by the following force fields: DREIDING\cite{Mayo1990}, GAFF\cite{Wang2004} and GAFF2\cite{He2020}. For constructing the polymer membrane, we have adopted a procedure referred to as "annealing" which is shown in Table \ref{suptable1}. We have built all polymer membranes with a fixed chain length of 30 monomers. In Fig.\ref{supfig4}, we show CPDMD simulation results for membranes having a thickness of 6 nm and 8 nm, respectively. Overall, the CO2-permeability does not depend on the choice of force fields or the polymeric membrane thickness. While the CO$_2$-permeability values obtained are similar for the polymers analyzed here, we observe some variability due to the amorphous nature of polymer membranes which is more pronounced for the thicker membranes. Based on the results, we chose the DREIDING force field because it allows us to cover a broader range of polymers. For the purpose of the main study and to account for computational resources, the membrane thickness was set to 6 nm.

In a second step, we have investigated the the equilibration process for obtaining the polymer membrane and the chain length which is shown in Fig.\ref{supfig5}. Specifically, we have considered two different equilibration processes referred to as "compression" and "annealing" which are shown in Table \ref{suptable1}. The polymer chain length is limited by the number heavy atoms, i.e. atoms other than hydrogen, and we have chosen the following values: 200, 500, and 800. In Fig.\ref{supfig5}, we show the CO$_2$-permeabilities obtained for all polymers with the different chain lengths and equilibration processes used. By comparing the data, we obtain best results in "compression" equilibration with a chain length of 800 heavy atoms. 

To confirm the choice of parameters, we have performed for six of the homo-polymers shown in Fig.\ref{supfig3} a set of five CPDMD simulations each. The results are shown in Fig.\ref{supfig6}. Overall, we obtain reasonable agreement with literature values for BZ-CF3, IBPA, PIM-PI-EA and PEO, despite the large error bars for BZ-CF3 and PEO. The simulated CO$_2$-permeabilities of TDA1-DM and PI-5 are higher than the literature values, however, one of the PI-5 samples is close to the experimental value. We note that due to the amorphous nature of polymers, both experimental and simulations results typically exhibit large error bars\cite{Kong2019}.

In a final step, we have determined the stop criterion for the CPDMD simulations. In Fig.\ref{supfig7}a, the vertical dashed lines indicate the times at which the polymer membranes reach the saturation level and, therefore, steady-state filtration. In other words, the number of CO$_2$ molecules inside the membrane as a function of time reaches a plateau. Similarly, the permeability curves in Fig.\ref{supfig7}b do not show significant change past that point in time which means that the time to stop the simulation is reached. To further exemplify the process, the evolution of the CO$_2$ density profile across a polymer membrane is shown in Fig.\ref{supfig8} for one of the top-three ranked polymers generated by the AI method.

\clearpage
\begin{table}
  \includegraphics[width=\linewidth]{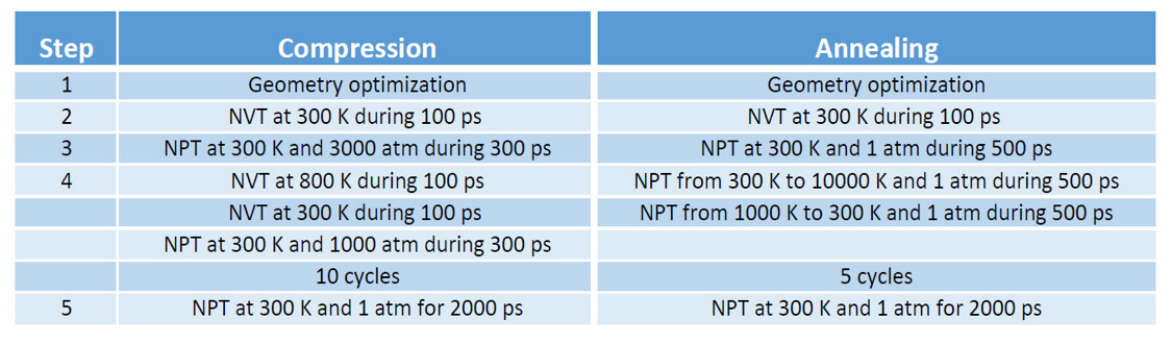}
  \caption{Comparison of two equilibration processes for obtaining the polymer membrane morphology.}
  \label{suptable1} 
\end{table}

\begin{figure}[h]
  \includegraphics[width=\linewidth]{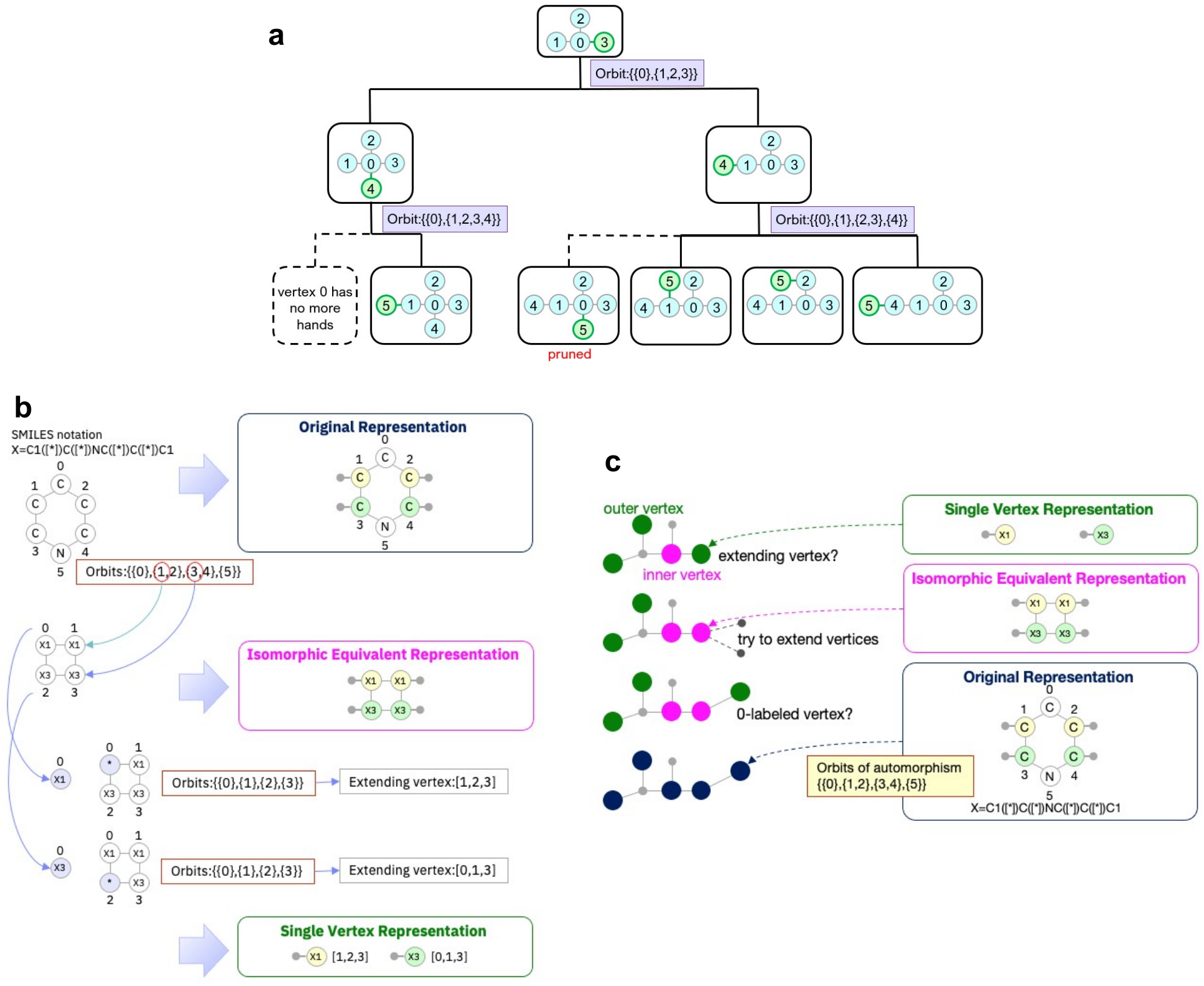}
  \caption{(a) Schematic visualization of the Molecular-Customized McKay’s Canonical Construction Path algorithm. Visual conceptions of the advanced versions of the Molecular-Customized McKay’s Canonical Construction Path algorithm with (b) sub-structure representations and (c) molecular construction example, respectively. Three level of sub-structures with graph representations are considered: single vertex representation for canonical construction path check, isomorphic equivalent representation for extending vertex check, and original representation for counting fragment occurrences.}
  \label{supfig2}
\end{figure}

\begin{figure}[h]
  \includegraphics[width=\linewidth]{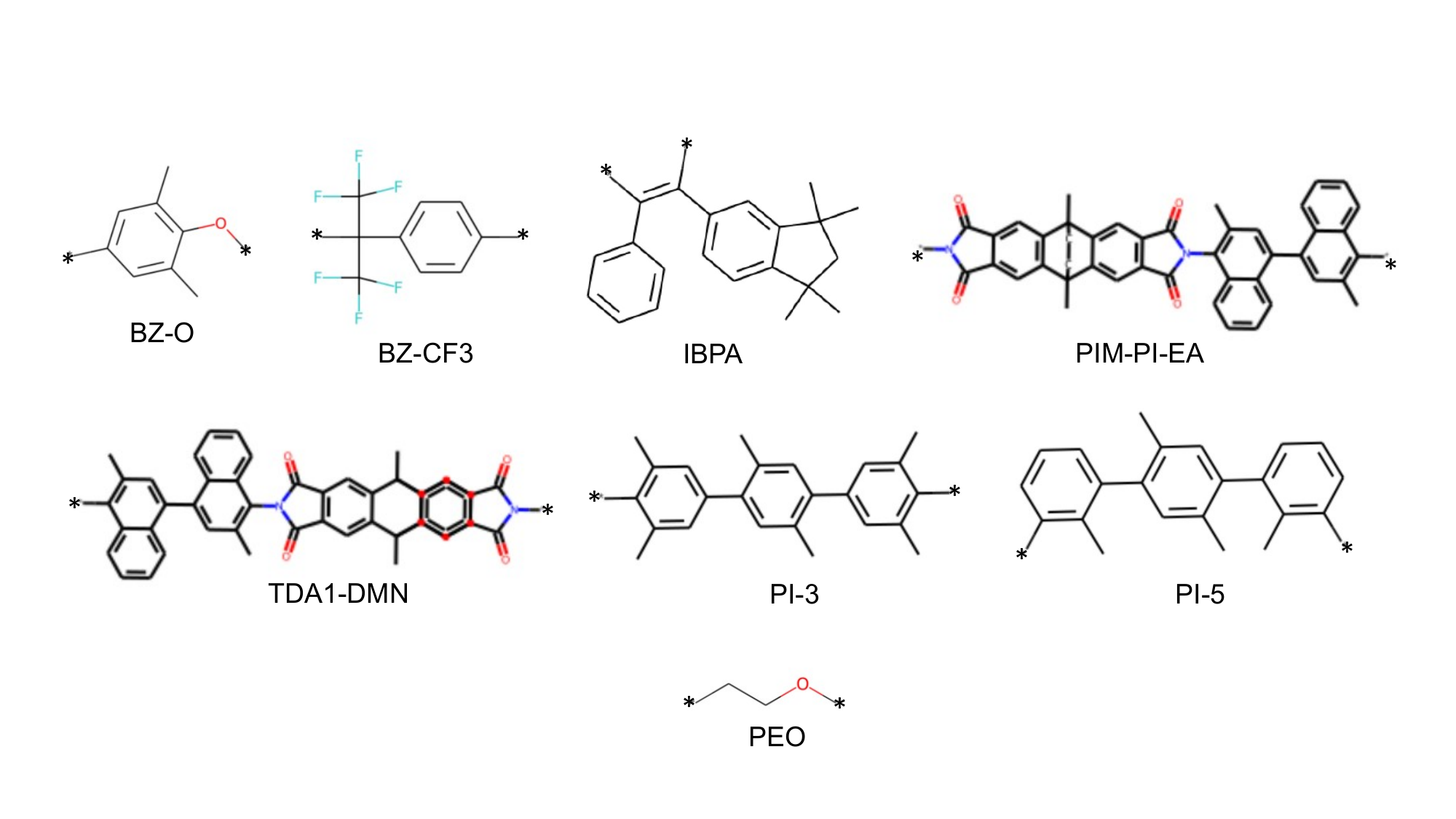}
  \caption{Representative monomer units chosen for Constant Pressure Difference Molecular Dynamics simulations of polymers. Experimental permeability values for benchmarking purpose were obtained from the literature: BZ-O \cite{Powell2006}, BZ-CF3\cite{Powell2006}, IBPA\cite{Yampolskii2012}, PIM-PI-EA\cite{Rogan2014}, TDA1-DMN\cite{Ghanem2016}, PI-3\cite{Powell2006} , PI-5\cite{Powell2006} and PEO\cite{Powell2006}.}
  \label{supfig3}
\end{figure}

\begin{figure}[h]
\centering
  \includegraphics[width=\linewidth]{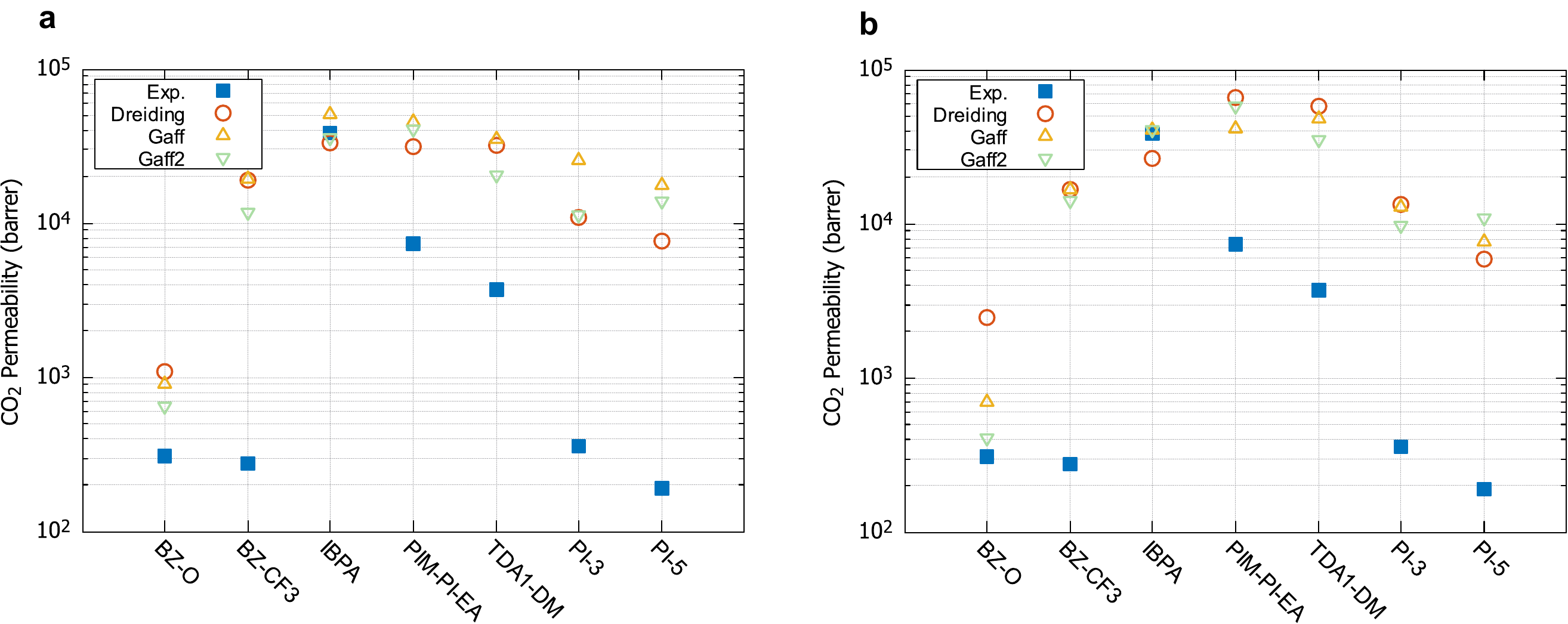}
  \caption{Effect of choice of force field on simulated CO$_2$-permeability for polymer membranes having a thickness of (a) 6nm and (b) 8 nm, respectively. }
  \label{supfig4}
\end{figure}

\begin{figure}[h]
\centering
\includegraphics[width=\linewidth]{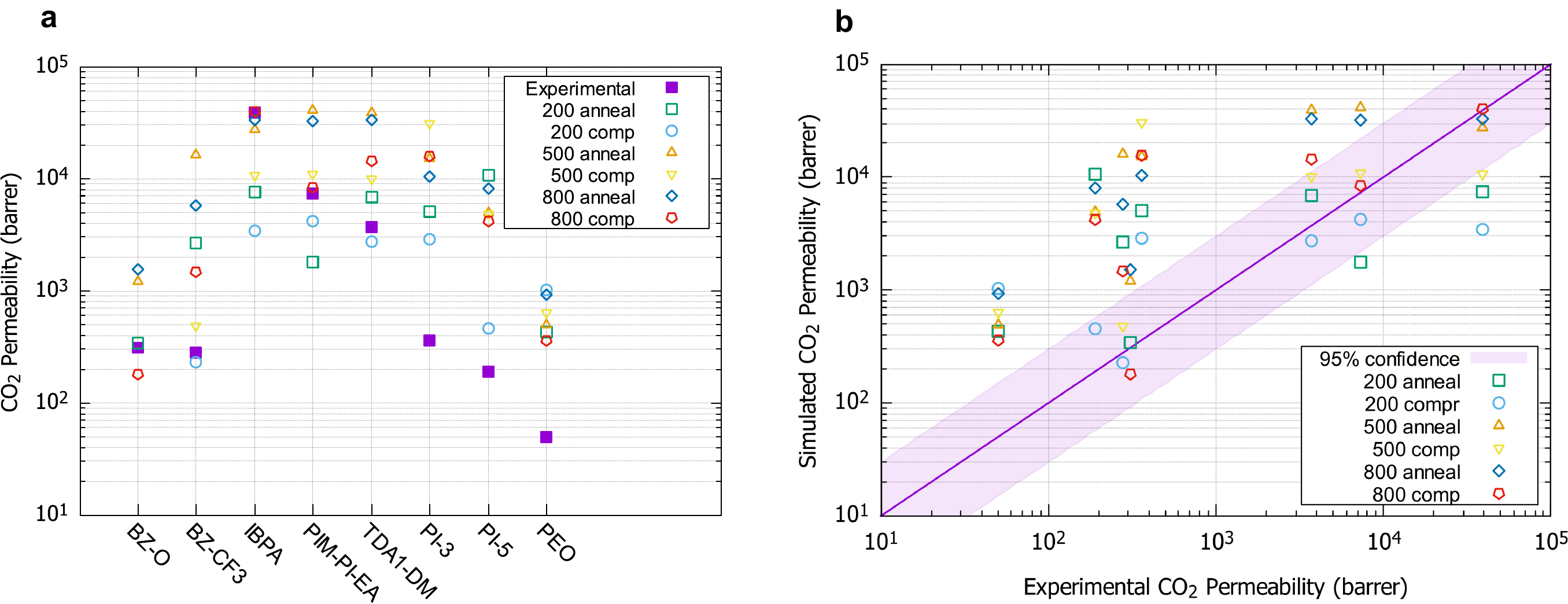} 
  \caption{Effect of choice of polymer chain length and membrane equilibration process. (a) CO$_2$-permeability for representative monomer units. (b) Simulated CO$_2$-permeability versus experimental CO2 permeability for representative monomer units.}
  \label{supfig5}
\end{figure}

\begin{figure}[h]
\centering
  \includegraphics[width=10cm]{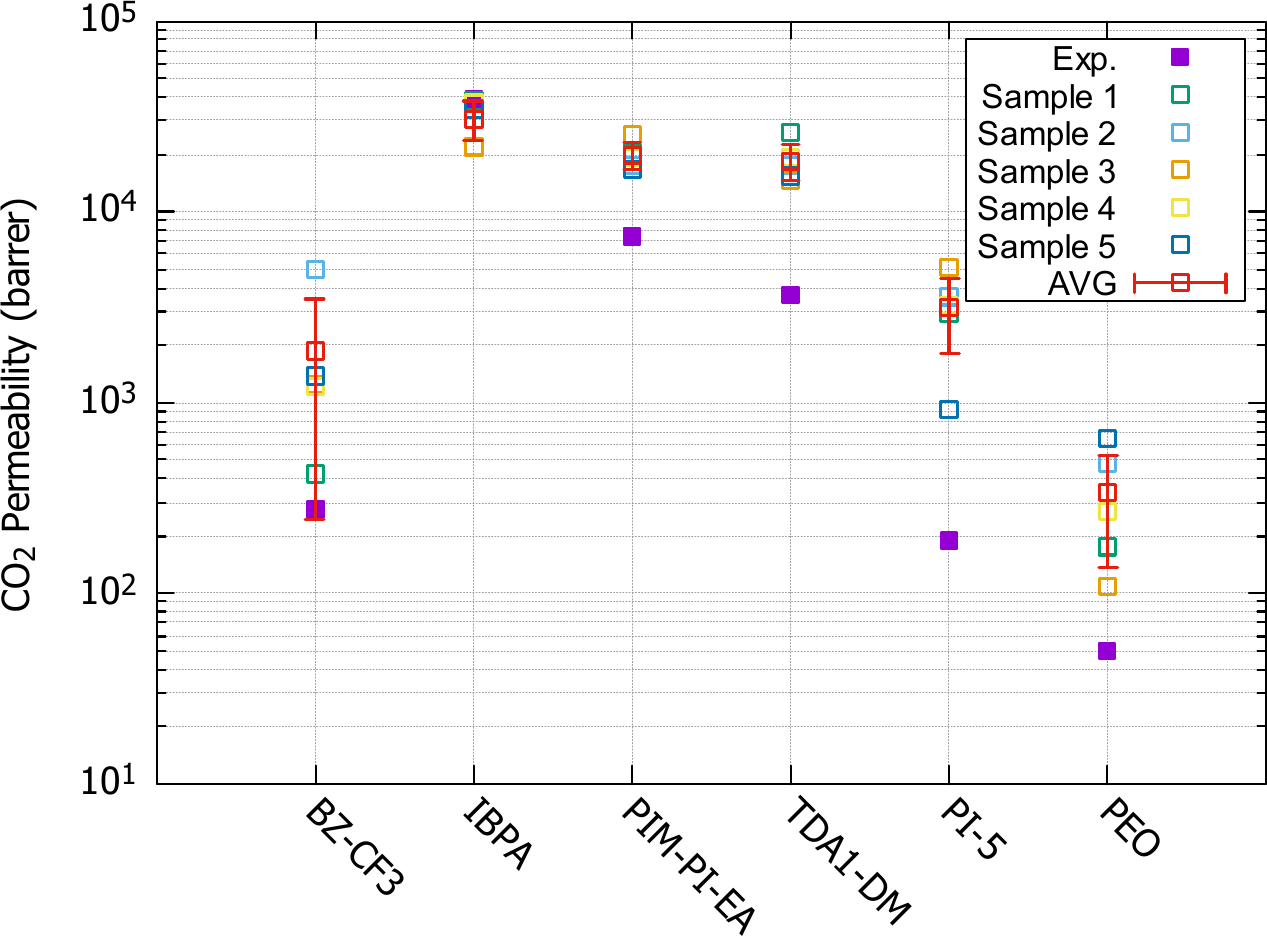}
  \caption{Constant Pressure Difference Molecular Dynamics (CPDMD) benchmark. For each representative monomer, the simulated CO$_2$-permeability values obtained for five polymer sample representations are plotted as open squares. The average permeability values obtained for each polymer are plotted as red squares. Experimental values obtained from the literature are plotted as solid squares: BZ-CF3 from \cite{Powell2006}, IBPA\cite{Yampolskii2012}, PIM-PI-EA\cite{Rogan2014}, TDA1-DMN\cite{Ghanem2016}, PI-5\cite{Powell2006} and PEO\cite{Powell2006} .}
  \label{supfig6}
\end{figure}

\begin{figure}[h]
  \centering
  \includegraphics[width=\linewidth]{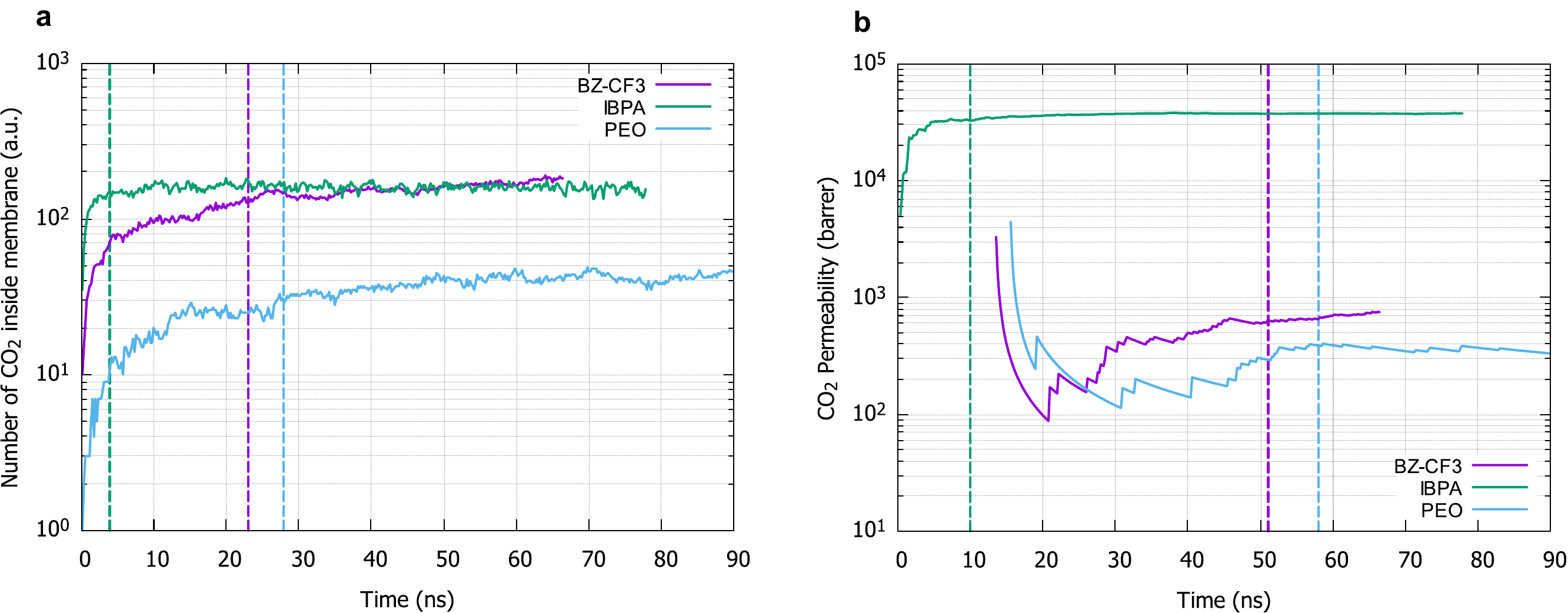}
  \caption{Constant Pressure Difference Molecular Dynamics (CPDMD) simulations of polymer membrane filtration. (a) Number of CO$_2$ molecules inside the polymer membrane. (b) CO$_2$-permeability of the polymer membrane as function of time. The vertical dashed lines in (a) indicate the times at which the simulations reach a steady state: 4 ns, 23 ns, and 28 ns for IBPA, BZ-CF3, and PEO, respectively. The vertical dashed lines in (b) indicate the time at which the CO$_2$-permeability reach steady state; 10 ns, 51 ns, and 58 ns for IBPA, BZ-CF3 and PEO, respectively.}
  \label{supfig7}
\end{figure}

\begin{figure}[h]
  \centering
  \includegraphics[width=10cm]{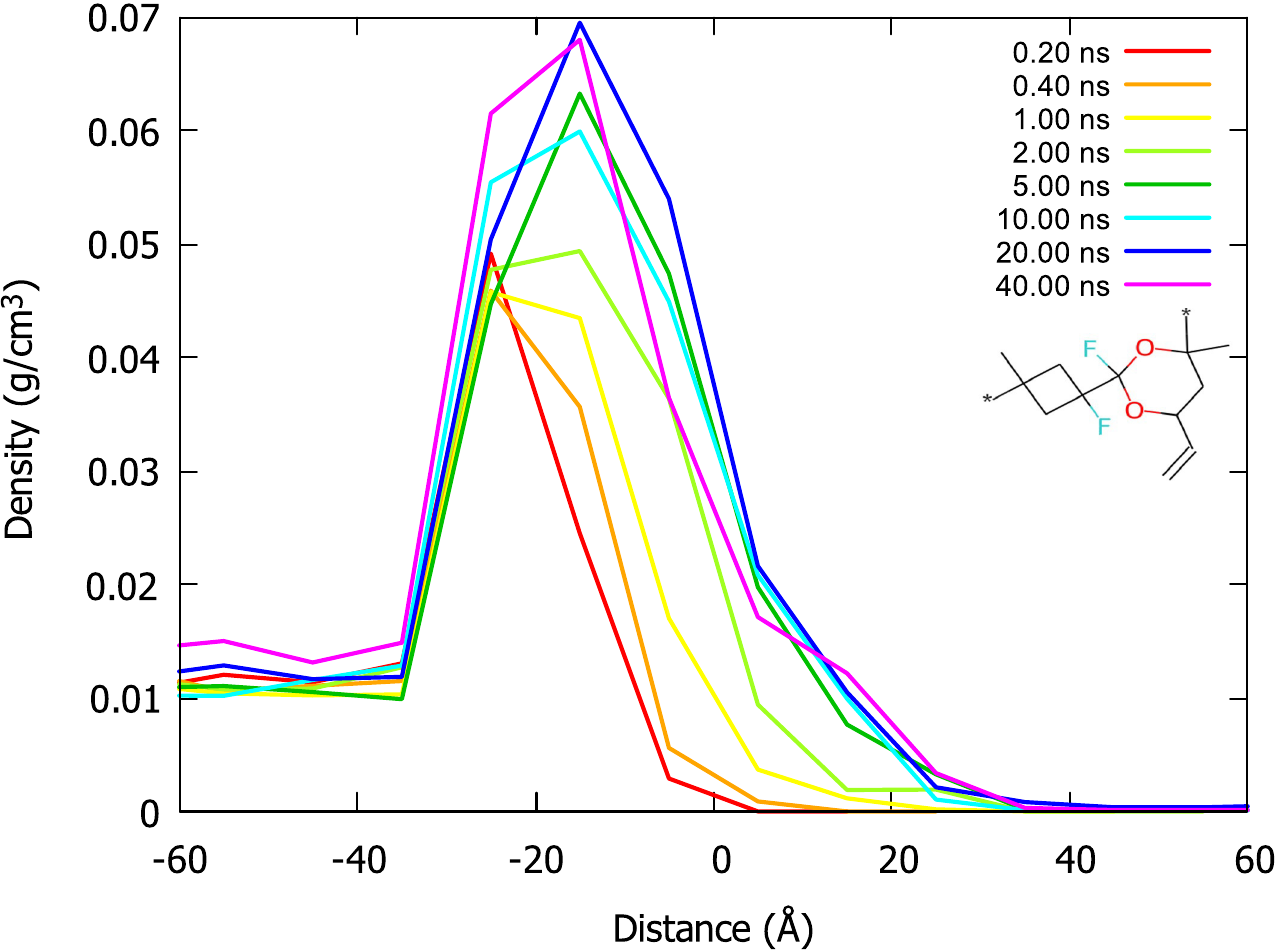}
  \caption{CO$_2$ density profile across a polymer membrane simulated by using a representative, AI discovered monomer unit.}
  \label{supfig8}
\end{figure}

\clearpage
\textbf{References}


\end{document}